\documentclass{emulateapj}
\usepackage{aas_macros}
\usepackage{epstopdf}
\usepackage{epsfig}
\usepackage{natbib}
\usepackage{float}
\usepackage{gensymb}
\usepackage{amsmath}
\usepackage{lipsum}
\usepackage{times}
\usepackage{comment}
\usepackage{color}
\usepackage{multirow}
\usepackage{soul}
\usepackage{CJKutf8}            

\newcommand{\Huanian}[1]{{\bf\color{red} HZ: #1}}
\definecolor{purple}{RGB}{160,32,240}

\definecolor{purple2}{RGB}{120,72,240}

\begin{document}

\title{Observing the Effects of Galaxy Interactions on the Circumgalactic Medium}

\author{Huanian Zhang \begin{CJK*}{UTF8}{gkai}(张华年)\altaffilmark{1}, Taotao Fang  (方陶陶) \altaffilmark{2}, Dennis Zaritsky\altaffilmark{1},  Peter Behroozi \altaffilmark{1},  Jessica Werk \altaffilmark{3}, and Xiaohu Yang  (杨小虎) \end{CJK*}\altaffilmark{4,5}}
\altaffiltext{1}{Steward Observatory, University of Arizona, Tucson, AZ 85719, USA; fantasyzhn@email.arizona.edu}
\altaffiltext{2}{Department of Astronomy, Xiamen University, Xiamen, Fujian, China }
  \altaffiltext{3}{Department of Astronomy, University of Washington, Seattle, WA 98195,  USA}
\altaffiltext{4}{Department of Astronomy, School of Physics and Astronomy, Shanghai JiaoTong University, Shanghai, 200240, China}
\altaffiltext{5}{Tsung-Dao Lee Institute, and Shanghai Key Laboratory for Particle Physics and Cosmology, Shanghai Jiao Tong University 
  Shanghai, 200240, China}

\begin{abstract}
We continue our empirical study of the emission line flux originating in the cool ($T\sim10^4$ K) gas that populates the halos of galaxies and their environments.  Specifically, we present results obtained for a sample of galaxy
pairs with a range of projected separations, {\bf $10 < {S_p/\rm kpc} < 200$}, and mass ratios $<$ 1:5, intersected by 5,443 SDSS lines of sight at projected radii of 10 to 50 kpc from either or both of the two galaxies. We find significant enhancement in H$\alpha$ emission and a  moderate enhancement in [N {\small II}]6583 emission for low mass pairs (mean stellar mass per galaxy, $\overline{\rm M}_*,  <10^{10.4} {\rm M}_\odot$) relative to the results from a control sample.
 This enhanced H$\alpha$ emission  comes almost entirely from sight lines located between the galaxies, consistent with a short-term, interaction-driven origin for the enhancement. 
 We find no enhancement in H$\alpha$ emission, but significant enhancement in [N {\small II}]6583 emission for high mass ($\overline{\rm M}_* >10^{10.4}{\rm M}_\odot$) pairs.  Furthermore, we find a dependence of the emission line properties on the galaxy pair mass ratio such that those with a mass ratio below 1:2.5 have enhanced [N {\small II}]6583 and those with a mass ratio between 1:2.5 and 1:5 do not. In all cases, departures from the control sample are only detected for close pairs ($S_p <$ 100 kpc).  Attributing an elevated [N {\small II}]6583/H$\alpha$ ratio to shocks, we infer that shocks play a role in determining the CGM properties for close pairs that are among the more massive and have mass ratios closer to 1:1.

\end{abstract}


\section{Introduction}

 Even before we developed an understanding of the central role that galaxy interactions, mergers, and accretion events play in the current picture of hierarchical galaxy evolution, such phenomena were a major area of study \citep[e.g.,][]{toomre}. A large set of literature now addresses how these events affect the central galaxy, including evidence for large inflows of gas into the centers of  galaxies \citep[e.g.,][]{Barnes1996, Rampazzo2005} or active galactic nuclei \citep[e.g.,][]{Springel2005,Hopkins2006}, mixing of gas as evidenced by the change of behavior in metallicity gradients \citep[e.g.,][]{Kewley2006,Kewley2010}, the enhancement of star formation and deficit in gas phase metallicity in close galaxy pairs \citep{Ellison2013}, and the prevalence of interaction signatures among post-starburst galaxies \citep{zabludoff}. If dynamical effects have measurable effects on the central galaxy, then they must have even larger effects on matter at the periphery.
 
Galaxies are surrounded by extended and diffuse gas, referred to as the circumgalactic medium (CGM), which is a critical but incompletely understood part of galactic ecosystems.
The CGM provides fuel for subsequent star formation activities \citep{spitzer} and serves as the dumping ground for galactic recycling and feedback \citep{CGM2017}. Although the  existence of this component was established 
from the study of absorption lines in the spectra of bright background objects \citep{bahcall,boksenberg,tytler} and continues to be an active field of research \citep[e.g.][]{steidel2010,menard2011,bordoloi2011,zhu2013a,zhu2013b,Werk2014,werk16,croft2016,croft2018,prochaska2017,Cai2017,Chen2017a, Chen2017b, Zahedy2017, lan2018,joshi2018, Johnson2018, Chen2019, Zahedy2019}, the nature of those studies, with a single line of sight through the halo of a modest number of galaxies, makes it difficult to explore the CGM properties in subclasses of galaxies. Here, specifically, we are interested in exploring how the CGM might be affected by galaxy-galaxy interactions. Simulations indeed predict that major mergers may have a dramatic effect on the metal content and ionization state of the gas in the CGM \citep{Hani2018}.

With the advent of the detection of optical emission lines from the CGM of normal, low redshift galaxies
\citep{zhang2016}, we now have a way to 
build up measurements of the CGM in specific sub-samples of galaxies and constrain the CGM properties as a function of many different variables. 
By suitably selecting subsamples drawn from 
the 
over 7 million lines of sight from the Sloan Digital Sky Survey \citep[SDSS DR12;][] {SDSS12}, 
we have 
characterized the nature of the emission lines and the CGM in low redshift galaxies and their environments \citep
{zhang2016,zhang2018a,Zhang2018b,Zhang2019a,Zhang2020}. We will now simply examine the corresponding emission line properties of lines of sight projected near close galaxy pairs. 
We adopt an imprecise $\Lambda$CDM cosmology with parameters
$\Omega_m$ = 0.3, $\Omega_\Lambda =$ 0.7, $\Omega_k$ = 0 and the dimensionless Hubble constant $h = $ 0.7 \citep[cf.][]{riess,Planck2018}.

\section{The Data}

We begin by describing the set of  paired galaxies about which we measure the emission flux from the cool CGM. 
We identify pairs using the  galaxy positions and redshfits from the Sloan  Digital  Sky  Survey  Data  Releases 7 catalog \citep[SDSS DR7]{Abazajian2009} and extract measures of each galaxy's
Sersic index ($n$) 
and absolute magnitude ($M$) from \citet{simard}, stellar mass (M$_*$) from \citet{Kauffmann2003a,Kauffmann2003b} and \citet{Gallazzi}, and current star formation rate (SFR) from \citet{Brinchmann}. For what we call our ``main" sample, we require the galaxy pair projected separation, $S_p$, between the two galaxies,  calculated from the average angular diameter distance of the two galaxies, to be greater than 10 kpc and less than 200 kpc, the velocity offset, $\Delta V$, to be less than 1000 km s$^{-1}$, and the mass ratio of the two galaxies to be no greater than 1:5 (to be clear, closer to 1:1 is allowed). In total, we find 52,721  galaxy pairs. Because we are interested in measuring the effect of the close interaction on the CGM properties, we also construct a ``control" sample of galaxy pairs for which  $400 < S_p/{\rm kpc} <$ 600,  $|\Delta V| < 1000$ km sec$^{-1}$ and the mass ratio less than 1:5. The control sample contains 75,007 pairs. We will also discuss the effect on our results of adopting a smaller velocity difference criterion of 500 km s$^{-1}$ below, but unless otherwise noted our discussion is based on the original criteria. To address the possibility that one galaxy may pair with different galaxies at different $S_p$, we keep only the pairing with the smallest $S_p$. This choice is informed by  our hypothesis that it is the closer pairings that have the greatest effect on the CGM. Of course, the smallest $S_p$ does not guarantee the smallest real separation, but this is the best we can do.

According both to previous theoretical studies \cite[e.g.,][]{keres} and the empirical study of the CGM emission line ratios \citep{Zhang2018b}, there is 
a qualitative change in the nature of the CGM for galaxies above and below a stellar mass of $\sim 10^{10.4}$ M$_\odot$. Therefore,
we divide our sample into a low mass subsample with $9.5<\log(\overline {\rm M}_*/{\rm M}_\odot)\le10.4$ and a high mass subsample with $10.4<\log(\overline {\rm M}_*/{\rm M}_\odot)\le11.5$, where $\overline{\rm M}_*$ is the average stellar mass of the galaxies in the pair system.

We collect the spectra for lines-of-sight projected within 50 kpc of either galaxy of each galaxy pair from the SDSS DR12. There are 5,443 such lines of sight for the main sample and 10,427 for total for control sample. 
For each such spectrum, we fit and subtract a 10th order polynomial to a 300 \AA\ wide section surrounding the wavelength of H$\alpha$ at the related galaxy redshift to remove the continuum. 
We then measure the residual H$\alpha$ and [N {\small II}]6583 flux within a velocity window corresponding to $\pm$275 km s$^{-1}$ from the related galaxy to capture the majority of the emission flux from the gas surrounding that galaxy. Of the two [N  {\small II}] emission lines near H$\alpha$, we measure only [N  {\small II}]6583 because, as we have determined previously \citep{zhang2016,zhang2018a,Zhang2018b}, [N {\small II}]6548 is far weaker and we have trouble measuring it in our stacks.
Hereafter, we refer to [N {\small II}]6583 simply as [N {\small II}]. 
This procedure replicates what we have done previously \citep{zhang2016} and various tests and controls are discussed in our previous studies.   We will also explore the sensitivity of our results to opening up the velocity window to $\pm$ 450 km s$^{-1}$.

\section{Results}

We present stacked measurements of the H$\alpha$ and [N {\small II}] fluxes for projected  radii between 10 and 50 kpc from each of the galaxies 
as a function of projected  pair separation for each of the two mass-selected subsamples.  The results\footnote{
The conversion factor to units between the values we present,  $10^{-17}$ erg cm$^{-2}$ s$^{-1}$ \AA$^{-1}$ and those used commonly in the literature to describe diffuse line emission, erg cm$^{-2}$ s$^{-1}$ arcsec$^{-2}$, is 1.7.} are presented in Table \ref{tab:data}.
In figures and tables we present the median emission line fluxes and the associated  uncertainties, estimated using a jackknife method.  Specifically, we randomly select half of the individual spectra, calculate the mean emission line flux, and repeat the process 1000 times to establish the distribution of measurements from which we quote the values corresponding to the 16.5 and 83.5 percentiles as the lower and upper uncertainties, respectively. We only present results for H$\alpha$ and [N {\small II}]  although 
we attempted to measure the flux for [O {\small II}]3727,3729 and  [O {\small III}]5007. We did not obtain measurements that were significantly greater than zero for any of the oxygen lines.

\begin{deluxetable*}{lrrrrr}
\tablewidth{0pt}
\tablecaption{The H$\alpha$ and [N {II}] fluxes vs. $S_p$ for different  galaxy pair samples}
\tablehead{
\colhead{Sample}&\colhead{Line}&\colhead{$f({\overline S}_p = 32$ kpc)}&\colhead{ $f({\overline S}_p = 79$ kpc)} & \colhead{$f({\overline S}_p = $ 154 kpc)} & \colhead{$f({\overline S}_p = $ 487 kpc)} \\ 
&&\multicolumn{4}{c}{[$10^{-17}$ erg cm$^{-2}$ s$^{-1}$ \AA$^{-1}$]} 
}
\startdata
\\
low mass  &H$\alpha$ & $0.033\pm0.005$  & $0.0044\pm0.0033$ & $-0.0001\pm0.0023$  & $0.0053\pm0.0014$ \\
&[N II]& $0.0069\pm0.0059$  & $0.0011\pm0.0033$ & $-0.0039\pm0.0024$ & $-0.0010 \pm 0.0016$\\
\\
high mass &H$\alpha$ & $-0.0033\pm0.0073$ & $0.0010\pm0.0038$ & $-0.0013\pm0.0027$ & $0.0014 \pm 0.0017$ \\
&[N II] & $0.015\pm0.007$  & $0.019\pm0.005$ & $-0.0039\pm0.0031$ & $0.0048 \pm 0.0018$ \\
\\
low mass ratio &H$\alpha$ & $0.016\pm0.007$ & $-0.0009\pm0.0035$ & $-0.0021 \pm 0.0020$ & $0.0011\pm0.0020$ \\
&[N II] & $0.015 \pm 0.006$ & $0.010 \pm 0.003$ & $-0.0045 \pm 0.0022$ & $0.0012 \pm 0.0014$ \\
\\
high mass ratio & H$\alpha$ & $0.023 \pm 0.007$ & $0.008 \pm 0.004$ & $0.0016 \pm 0.0027$ & $0.0034\pm0.0022$ \\
&[N II]& $0.0025 \pm 0.0083$ & $0.007 \pm 0.004$ & $-0.0015 \pm 0.0030$ & $0.0003\pm 0.0016$\\
\enddata
\label{tab:data}
\end{deluxetable*}

\begin{deluxetable*}{lrrrrrr}
\tablewidth{0pt}
\tablecaption{The H$\alpha$ and [N {II}] fluxes for the control and parameter-matched control samples}
\tablehead{
\colhead{Sample}& \colhead{Line}&\colhead{$f({\overline S}_p = $ 487 kpc)}  & \colhead{NN matched} & \colhead{mass matched} & \colhead{sSFR matched}  &  \colhead{All matched} \\ 
&&\multicolumn{5}{c}{[$10^{-17}$ erg cm$^{-2}$ s$^{-1}$ \AA$^{-1}$]} 
}
\startdata
\\
low mass  &H$\alpha$ &  $0.0053\pm0.0014$ &  $0.0052\pm0.0016$ & 0.0053 $\pm$ 0.0016 & 0.0046 $\pm$ 0.0018 & 0.0056 $\pm$ 0.0024 \\
&[N II]&  $-0.0010 \pm 0.0016$ &  $-0.0011\pm0.0016$ & $-$0.0018 $\pm$ 0.0016 & $-$0.0048 $\pm$ 0.0018 & $-$0.0037 $\pm$ 0.0025\\
\\
high mass &H$\alpha$  & $0.0014 \pm 0.0017$ &  $0.0029\pm0.0019$ & 0.0012 $\pm$ 0.0018 & 0.0007 $\pm$ 0.0020 & 0.0044 $\pm$ 0.0024\\
&[N II] & $0.0048 \pm 0.0018$ &  $0.0054\pm0.0020$ & 0.0046 $\pm$ 0.0019 & 0.0032 $\pm$ 0.0020 & 0.0031 $\pm$ 0.0026\\
\\
low mass ratio &H$\alpha$ &  $0.0011\pm0.0020$ & $0.0024\pm0.0015$ & 0.0011 $\pm$ 0.0014 & $-$0.0007 $\pm$ 0.0016 & $-$0.0011 $\pm$ 0.0022
\\
&[N II] & $0.0012 \pm 0.0014$ & $0.0004\pm0.0015$ & 0.0004 $\pm$ 0.0015 & 0.0002 $\pm$ 0.0017 &  $-$0.0006 $\pm$ 0.0021\\
\\
high mass ratio & H$\alpha$ &  $0.0034\pm0.0022$ & $0.0029\pm0.0017$ & 0.0036 $\pm$ 0.0016 & 0.0044 $\pm$ 0.0019 &  0.0010 $\pm$ 0.0022\\
&[N II]&  $0.0003\pm 0.0016$ & $-0.0002\pm0.0017$ & $-$0.0003 $\pm$ 0.0017 & $-$0.0044 $\pm$ 0.0019 &  $-$0.0008 $\pm$ 0.0025\\

\enddata
\label{tab:matched}
\end{deluxetable*}
\subsection{Emission Line Fluxes and Mean Stellar Mass}

\begin{figure}[htp]
\begin{center}
\includegraphics[width = 0.48 \textwidth]{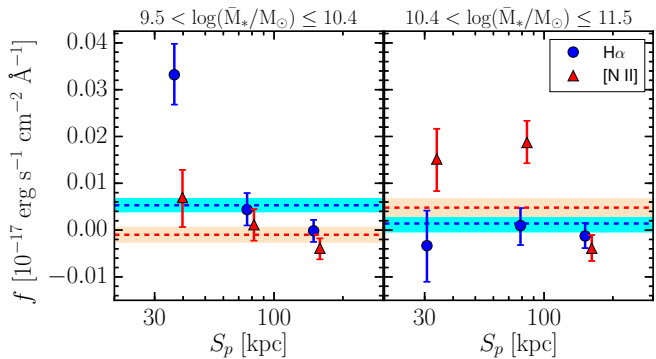}
\end{center}
\caption{The H$\alpha$ and [N II] emission fluxes as a function of projected galaxy pair separation, $S_p$, for low mass (left) and high mass (right) pairs.  
For visualization, we apply slight horizontal offsets between the H$\alpha$ and [N II] fluxes although they are all measured at the same $S_p$. The blue (red) dashed line represents the H$\alpha$ ([N II]) emission flux of the control sample. } 
\label{fig:binary}
\end{figure}

In Figure \ref{fig:binary} we present the H$\alpha$ and [N {\small II}] emission fluxes as a function of the projected separation between the two galaxies in the pair,  divided into two subsamples based on the mean stellar mass of the galaxy pair. Among the lower mass pairs, we find a significant ($>$ 4.7$\sigma$) enhancement of the H$\alpha$ flux over the control value at the smallest pair separations  and consistency with the control sample at greater separations. Among the higher mass pairs, the H$\alpha$ flux is not significantly enhanced at any pair separation.
Instead, [N {\small II}] is elevated in pairs with small and intermediate separations. Combining the inner two radial bins, the enhancement in [N {\small II}] is significant at the 97.5\% (2.5$\sigma{}$) confidence level. To determine the sensitivity of our measurements on our definition of pairs, we decreased the velocity difference criterion for pair identification to 500 km s$^{-1}$ and find only modest ($< 10$\%) flux differences.

Given the nature of galaxy interactions, and the possibility of high velocity gas inflows or outflows, we now examine the sensitivity of our results to our defined spectral velocity window. We expand the window to $\pm 450$ km s$^{-1}$, a compromise between expanding the window as wide as possible and minimizing the introduction of additional noise and contamination. We find a diminution of the H$\alpha$ signal for low mass pairs at the smallest $S_p$, from $0.033 \pm 0.006$ to $0.017 \pm 0.005$ (a 2$\sigma$ difference), which is still consistent with a significant flux enhancement but suggestive that there is no additional large flux contribution from high velocity gas that we are missing in our original measurement. The differences obtained in H$\alpha$ fluxes for either the next $S_p$ bin and for the control bin are statistically insignificant, as are the results for the [N II] detection in the high mass pair subsample. We conclude that the results are robust to plausible changes in the velocity window.

In interpreting our results, we are  concerned  that  our selection of  pairs  of different projected separation is somehow selecting pairs in different environments. We have previously found that the CGM emission line properties are affected by environment \citep{Zhang2019a}, and therefore, an environmental difference among our various subsamples could affect what we observe here.  One measure of environment that we have used previously is the number of nearby neighbors \citep[NN; see][for details]{Zhang2019a}. 

We find small, but significant differences in the mean NN value of the main and control samples,  whether we consider the low or high mass subsamples. Furthermore, we find that the distributions of NN are statistically different with greater than 99\% confidence using a KS test. Although the differences are small, they are a cause of concern. To address whether these differences are creating corresponding differences in the resulting control sample line fluxes, we create control samples that match the NN distribution of the comparison sample by randomly drawing an NN-matched distribution from the full set of control sample lines-of-sight. We present the results in Table \ref{tab:matched} and conclude that the differences in NN distributions are not responsible for the detected signal.

Similarly, we can test whether differences in the distributions of other parameters between the control and comparison samples might be affecting our result. We consider the distributions of S\'ersic index, $n$, to represent morphology, $r_p$, M$_*$, and specific star formation rate, sSFR. Using a KS test to identify which parameters might present significant differences in their distributions, we find statistically significant differences only for M$_*$ and sSFR. We perform the same exercise as we did for NN to create control samples that match the control sample parameter distribution to that of the comparison sample. In these cases, we match to the M$_*$ and sSFR distributions. The results of those calculations are also presented in Table \ref{tab:matched}. We find no significant ($> 2\sigma$) differences in the flux values derived from the original and matched control samples.

Finally, we present the results of   simultaneously matching to the distribution of sSFR, NN and M$_*$. To do this, we force the 3D distribution of  sSFR, NN and M$_*$ of the control sample to be the same as that of the data (results in rightmost column of Table \ref{tab:matched}). This test runs the risk of dividing the sample of sightlines too finely and, by doing so, depending too much on a limited set of data. Despite this concern, our results are consistent both with our original measurements for the control samples and with the 1-D distribution-matched results (again to within 2$\sigma$). We conclude that differences among the control and comparison samples, at least with regard to the limited parameter set explored, are not responsible for the results we feature here.

\subsection{Emission Line Fluxes and Mass Ratio}

\begin{figure}[htbp]
\begin{center}
\includegraphics[width = 0.48 \textwidth]{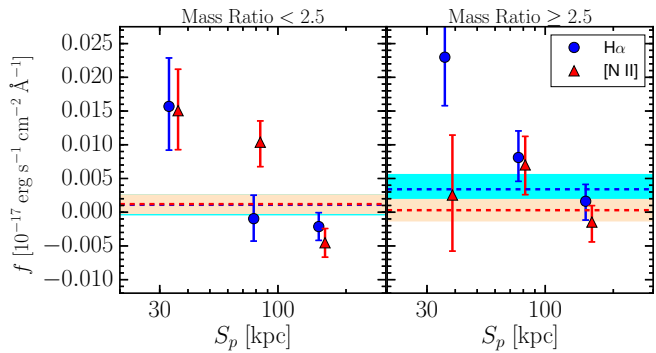}
\end{center}
\caption{The left panel presents the emission line flux measurements for the low mass ratio subsample as a function of  galaxy pair separation, while the right panel displays the same measurements for the high mass ratio subsample. The symbols are the same as in those in Figure  \ref{fig:binary}. }
\label{fig:binary_massratio}
\end{figure}

 Our selection of galaxy pairs imposed a mass ratio limit of 1:5, but within that range we now define a subsample that is closer to 1:1, where one might expect the effect of any interaction to be stronger. In Figure \ref{fig:binary_massratio},  
 we present results when we divide the sample by mass ratio rather than by mean stellar mass.
We find that H$\alpha$ is enhanced at small galaxy pair separations in both cases (2.5$\sigma$ and 2.8$\sigma$, low and high mass ratios, respectively), presumably because both mass ratio subsamples are dominated by the low mass pairs that showed H$\alpha$ enhancement at small $S_p$.
More interestingly, we find a significant enhancement of [N {\small II}] (3.5$\sigma$ significance when the inner two $S_p$ bins are combined) for pairs with mass ratios closer to 1:1. This enhancement is present, and even more significant among the low mass subsample of these pairs, which overall did not show such enhancement in [N {\small II}] (Figure \ref{fig:binary}). 

Again, using the KS test, we find slight, but significant, differences in the galaxy properties between main and control samples. As before, we construct parameter matched control samples to ascertain whether those differences are in any way responsible for the results we are finding. The fluxes derived from the parameter-matched samples are presented in Table \ref{tab:matched} and we conclude that the differences in the control samples are not responsible for our results.


\subsection{Dependence on Geometry}

We turn to investigating the dependence of the signal on the location  of the sight line relative to the pair system. We define the set of sight lines between the two galaxies of the pair to be those that lie within the smallest circle on the sky that intersects the two galaxies and refer to those as interior sight lines. The remaining sight lines outside of this circle are considered to be associated primarily with one of the two galaxies, and we refer to those as exterior sight lines. 

\begin{figure}[htbp]
\begin{center}
\includegraphics[width = 0.48 \textwidth]{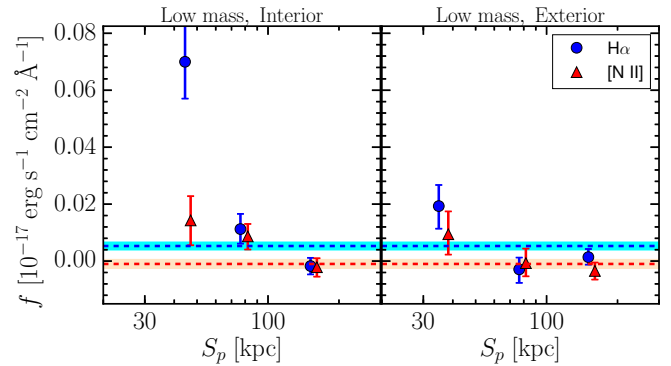}
\end{center}
\caption{The left panel presents the emission line flux measurements for the low mass pair sample using only  interior sight lines (left panel) and only exterior sight lines (right panel) as a function of  galaxy pair projected separation. The symbols are the same as in those in Figure  \ref{fig:binary}. }
\label{fig:binary_oren_lowmass}
\end{figure}

In Figure \ref{fig:binary_oren_lowmass} we display results obtained using only interior  sight lines (left panel) and only exterior 
sight lines (right panel) for the low mass pairs. The flux difference is stark and illustrates that the result shown in the left panel of Figure \ref{fig:binary} is contributed almost entirely by interior sight lines. Because the pair galaxies are not necessarily all on first approach, we interpret this result to indicate that the elevated H$\alpha$ emission is a short-lived phenomenon existing where the CGM from the two galaxies is actively interacting. 
We find no analogous difference among the elevated [N {\small II}] emission results when using interior and exterior sight lines for the high mass subsample. We interpret this result to indicate that this phenomenon is  long-lasting and affects the entire halo of each galaxy.

We find no significant differences in either the H$\alpha$ or [N {\small II}] fluxes between the the interior and exterior sight lines for the high mass sample. The lack of a difference in H$\alpha$ is not surprising because we did not detect any H$\alpha$ enhancement in the high mass sample to begin with. However, the lack of a difference for [N {\small II}] is interesting and may be signaling that the elevation in [N {\small II}] flux affects the entire halo of each galaxy in the pair.

\section{Discussion}

We now briefly discuss aspects of possible interpretations of the observational results.

\subsection{Physical and Projected Galaxy Pairs}

We use cosmological simulations to estimate the fraction of physical pairs in our samples. Specifically, we use simulations that utilize halo merger trees from the {\sl Bolshoi-Planck} simulation \citep{Klypin16,RP16}, with halos found using the \textsc{Rockstar} phase-space halo finder \citep{BehrooziRockstar},  merger trees generated with the \textsc{Consistent Trees} code \citep{BehrooziTrees}, and, finally, stellar masses modeled with the \textsc{UniverseMachine} code \citep{Behroozi2018}. 

When we apply our selection criteria for the main sample, we find that 44.6\% of the identified pair systems have 3-D separation, $R_{\rm 3D},  < 200$ kpc. 
Similarly, if we limit our selection to pairs with $S_p < 50$ kpc, we find that 43.6\% have $R_{\rm 3D} < 50$ kpc.
Both of these results suggest that our measured line emission enhancements are likely underestimated, perhaps by as a much as a factor of two, due to significant contamination by wider separation system.

\subsection{Outer Region Star Formation}

In the low mass pair sample, we find elevated H$\alpha$ fluxes in the innermost bin. This may be the result of changing CGM conditions in these interacting systems. For example, the interaction might have increased the density of the emitting gas and therefore increased the emission, which depends on density squared.  Simulations, which are used to inform the interpretation of observed absorption lines, find significantly higher H, C, and O covering fractions in the vicinity of merging galaxies compared to those of isolated galaxies \citep{Hani2018}, although model-dependent ionization effects can lead to decreases in the covering fraction of associated, commonly-observed ions (H {\small I}, C {\small IV}, O {\small VI}).  This complimentary result suggests a higher level of emission from the CGM in galaxy mergers, as we have found in the current study. 

On the other hand, interactions are often considered to give rise to increased star formation. This may even be true in the outer disks of galaxies. One spectacular example of outer disk star formation, M 83 \citep{thilker}, also appears to have interacted recently \citep{malin}. The outer disk star forming regions, at least in M 83, extend out to several tens of kpc.  In a sample of nearby HI-selected galaxies, \cite{werk10} find that $\sim$10\% show outlying HII regions in H$\alpha$ emission beyond 2 $\times$ the 25th magnitude R-band isophotal radius, often well beyond 20 kpc. Although our stacking methodology rejects lines-of-sight with anomalously high H$\alpha$ fluxes, some lines-of-sight through the diffuse outer disks of systems like M 83 could perhaps be responsible for the observed excess. Imaging of the emission lines in pair systems will allow us to determine if the excess comes from highly localized star formation regions or is due to a more general halo phenomenon.

\subsection{Signature of Shocks}

We have previously exploited line ratios to help us understand the nature of excitation mechanism of the lines we are observing \citep{Zhang2018b}. Here, we are particularly interested in the role shocks may be playing when the galaxies interact.  As discussed by \cite{Rich2010} and \cite{Farage2010}, shock models, for velocities $>$ 200 km s$^{-1}$, predict $\log$([N {\small II}]6583/H$\alpha$) $>$ 0.2 and $\log$([O {\small III}]/H$\beta$) $< 0$.
Of particular interest is that we find that $\log$([N {\small II}]6583)/H$\alpha$) increases in the interacting systems, and increases into a range of values consistent with shock excitation. The lower limit of  $\log$([N {\small II}]6583)/H$\alpha$) that we obtain for the high mass subsample is 0.65. 
This value is a lower limit because we calculate it using the upper bound on the H$\alpha$ flux (our measurement for the H$\alpha$ flux is formally negative, even though it is consistent with small positive values given the uncertainties). 
Because this value of the line ratio is well above the threshold of 0.2, we suggest that we may be viewing CGM shocks in interacting  galaxy pairs. 
Furthermore, we note that the increase in [N{\small II}] is most noticeable when the galaxies are comparable in mass (Figure \ref{fig:binary_massratio}), and so shocks presumably have comparable effects on the CGM surrounding both members.

This is not the first indication of shocks in the CGM of interacting galaxies. For example,  \cite{Yoshida2016} noted that the correspondence between the H$\alpha$-emitting nebula surrounding the starburst/merging galaxy NGC 6240 and the X-ray (0.7$–$1.1 keV) emitting hot gas  \citep{Nardini2013} suggests shock-heating as a way to excite both phases of the gas. Unfortunately, there are complicating factors in the interpretation of our result and sophisticated modeling will ultimately be needed. For example, $\log (\rm [N~ {\small II}]6583/H\alpha)$ is known to also be a metallicity indicator \citep{Pettini2004}. The simulations of \cite{Hani2018} indeed find that galaxy major mergers increase the CGM metallicity by 0.2$-$0.3 dex immediately following the interaction. As such, some of the variation we observe in this line ratio could also be due to either the introduction of low metallicity gas at small radii or the dredging up of higher metallicity gas from inner radii out into the halo.

\section{Summary}

We present measurements of H$\alpha$ and [N {\small II}] emission line flux as a function of projected distance for  galaxy pairs.  We define the sample by requiring that the galaxy pair projected distance between the two galaxies be less than 200 kpc, that the recessional velocity difference be less than 1000 km s$^{-1}$, and that the mass ratio be less than 1:5. Meanwhile, we construct a control sample where the only difference is that the galaxy pair projected distance range is from 400 to 600 kpc. We measure the H$\alpha$ and [N {\small II}] emission line flux within projected separations of 10 $<r_p/{\rm kpc} < $ 50 around either galaxy of the pair system. 

We identify a variety of differences in the CGM properties of galaxy pairs in our main and control samples. 
First, we find that for lower mass pairs (mean stellar mass of $10^{10.03}$ M$_\odot$), there is a significant ($> 4.7  \sigma$) enhancement in the H$\alpha$ emission line flux, relative to that in control sample, at small separations (mean projected separation, $\overline{S}_p, = 32$ kpc). At other separations less than 200 kpc the data and control sample are statistically consistent for both H$\alpha$ and [N {\small II}]. Second, among high mass pairs (mean stellar mass of $10^{10.68}$ M$_\odot$), [N {\small II}]  emission line flux is significantly ($> 2.5 \sigma$) enhanced {for $S_p < 100$ kpc.}
H$\alpha$ emission flux is consistent with the control value  at all radii. 

In results that further constrain possible interpretations, we also find that the mass ratio of the pair plays a role. We detect a statistically significant ($3.5\sigma$) enhancement of the [N {\small II}] emission line flux for the half of our pair sample with mass ratios closer to 1:1 and no such enhancement among the higher mass ratio pairs.  Furthermore, we find that the bulk of the H$\alpha$ flux enhancement at low projected separations for the low mass pair sample comes from lines of sight located between the two galaxies. We interpret this result to indicate that the elevated H$\alpha$ emission is a short-lived phenomenon existing where the CGM from the two galaxies is actively interacting. We find no analogous difference for the high mass subsample, which we interpret to indicate that in this case the phenomenon is  long-lasting and affects the entire halo of each galaxy.
 
The rise in H$\alpha$ could be attributed to a variety of factors, but we attribute the increase of [N {\small II}] to shocks. The preliminary indication is that the manifestation of shocks is strongest in the more massive systems and in those pairs where the mass ratio is closer to 1:1. We conclude that interactions begin to grossly affect the CGM at separations $\lesssim$ 100 kpc, which then, presumably, affects how that gas interacts with the central galaxy. Emission line maps of such systems would be of tremendous value in assessing interaction between the circumgalactic media of the interacting galaxies.

\section{Acknowledgments}

DZ and HZ acknowledge financial support from NSF grant AST-1311326.  T. Fang is supported by the National Key R\&D Program of China No. 2017YFA0402600, and NSFC grants No. 11525312, 11890692. JW acknowledges support from a 2018 Sloan Foundation Fellowship.  This work is supported by the 973 Program (Nos. 2015CB857002) and national science foundation of China (grant Nos. 11833005, 11890692, 11621303).  We thank the referee for their constructive suggestions. We also thank the support
of the Key Laboratory for Particle Physics, Astrophysics and
Cosmology, Ministry of Education.  The authors gratefully acknowledge  the SDSS III team for providing a valuable resource to the community.
Funding for SDSS-III has been provided by the Alfred P. Sloan Foundation, the Participating I institutions, the National Science Foundation, and the U.S. Department of Energy Office of Science. The SDSS-III web site is http://www.sdss3.org/.

SDSS-III is managed by the Astrophysical Research Consortium for the Participating Institutions of the SDSS-III Collaboration including the University of Arizona, the Brazilian Participation Group, Brookhaven National Laboratory, Carnegie Mellon University, University of Florida, the French Participation Group, the German Participation Group, Harvard University, the Instituto de Astrofisica de Canarias, the Michigan State/Notre Dame/JINA Participation Group, Johns Hopkins University, Lawrence Berkeley National Laboratory, Max Planck Institute for Astrophysics, Max Planck Institute for Extraterrestrial Physics, New Mexico State University, New York University, Ohio State University, Pennsylvania State University, University of Portsmouth, Princeton University, the Spanish Participation Group, University of Tokyo, University of Utah, Vanderbilt University, University of Virginia, University of Washington, and Yale University.

\bibliography{bibliography}

\begin{thebibliography}{62}
\expandafter\ifx\csname natexlab\endcsname\relax\def\natexlab#1{#1}\fi

\bibitem[{{Abazajian} {et~al}\mbox{.}(2009){Abazajian}
  {et~al.}}]{Abazajian2009}
{Abazajian} K.~N., {et~al.}, 2009, \apjs, 182, 543

\bibitem[{{Alam} {et~al}\mbox{.}(2015){Alam}, {Albareti}, {Allende Prieto},
  {Anders}, {Anderson}, {Anderton}, {Andrews}, \& et~al.}]{SDSS12}
{Alam} S., {Albareti} F.~D., {Allende Prieto} C., {Anders} F., {Anderson}
  S.~F., {Anderton} T., {Andrews} B.~H., et~al., 2015, \apjs, 219, 12

\bibitem[{{Bahcall} \& {Spitzer}(1969)}]{bahcall}
{Bahcall} J.~N., {Spitzer}, Lyman J., 1969, \apjl, 156, L63

\bibitem[{{Barnes} \& {Hernquist}(1996)}]{Barnes1996}
{Barnes} J.~E., {Hernquist} L., 1996, \apj, 471, 115

\bibitem[{{Behroozi} {et~al}\mbox{.}(2018){Behroozi}, {Wechsler}, {Hearin}, \&
  {Conroy}}]{Behroozi2018}
{Behroozi} P., {Wechsler} R., {Hearin} A., {Conroy} C., 2018, ArXiv e-prints

\bibitem[{{Behroozi}, {Wechsler} \& {Wu}(2013){Behroozi}, {Wechsler}, \&
  {Wu}}]{BehrooziRockstar}
{Behroozi} P.~S., {Wechsler} R.~H., {Wu} H.-Y., 2013, \apj, 762, 109

\bibitem[{{Behroozi} {et~al}\mbox{.}(2013){Behroozi}, {Wechsler}, {Wu},
  {Busha}, {Klypin}, \& {Primack}}]{BehrooziTrees}
{Behroozi} P.~S., {Wechsler} R.~H., {Wu} H.-Y., {Busha} M.~T., {Klypin} A.~A.,
  {Primack} J.~R., 2013, \apj, 763, 18

\bibitem[{{Boksenberg} \& {Sargent}(1978)}]{boksenberg}
{Boksenberg} A., {Sargent} W.~L.~W., 1978, \apj, 220, 42

\bibitem[{{Bordoloi} {et~al}\mbox{.}(2011){Bordoloi}, {Lilly}, {Knobel},
  {Bolzonella}, {Kampczyk}, {Carollo}, \& et~al}]{bordoloi2011}
{Bordoloi} R., {Lilly} S.~J., {Knobel} C., {Bolzonella} M., {Kampczyk} P.,
  {Carollo} C.~M., et~al, 2011, \apj, 743, 10

\bibitem[{{Brinchmann} {et~al}\mbox{.}(2004){Brinchmann}, {Charlot}, {White},
  {Tremonti}, {Kauffmann}, {Heckman}, \& {Brinkmann}}]{Brinchmann}
{Brinchmann} J., {Charlot} S., {White} S.~D.~M., {Tremonti} C., {Kauffmann} G.,
  {Heckman} T., {Brinkmann} J., 2004, MNRAS, 351, 1151

\bibitem[{{Cai} {et~al}\mbox{.}(2017){Cai}, {Fan}, {Yang}, {Bian}, {Prochaska},
  {Zabludoff}, {McGreer}, {Zheng}, {Green}, {Cantalupo}, {Frye}, {Hamden},
  {Jiang}, {Kashikawa}, \& {Wang}}]{Cai2017}
{Cai} Z. {et~al.}, 2017, \apj, 837, 71

\bibitem[{{Chen}(2017{\natexlab{a}})}]{Chen2017b}
{Chen} H.-W., 2017{\natexlab{a}}, Astrophysics and Space Science Library, Vol.
  434, {Outskirts of Distant Galaxies in Absorption}, {Knapen} J.~H., {Lee}
  J.~C., {Gil de Paz} A., eds., p. 291

\bibitem[{{Chen}(2017{\natexlab{b}})}]{Chen2017a}
{Chen} H.-W., 2017{\natexlab{b}}, Astrophysics and Space Science Library, Vol.
  430, {The Circumgalactic Medium in Massive Halos}, {Fox} A., {Dav{\'e}} R.,
  eds., p. 167

\bibitem[{{Chen} {et~al}\mbox{.}(2019){Chen}, {Boettcher}, {Johnson}, {Zahedy},
  {Rudie}, {Cooksey}, {Rauch}, \& {Mulchaey}}]{Chen2019}
{Chen} H.-W., {Boettcher} E., {Johnson} S.~D., {Zahedy} F.~S., {Rudie} G.~C.,
  {Cooksey} K.~L., {Rauch} M., {Mulchaey} J.~S., 2019, \apjl, 878, L33

\bibitem[{{Croft} {et~al}\mbox{.}(2018){Croft}, {Miralda-Escud{\'e}}, {Zheng},
  {Blomqvist}, \& {Pieri}}]{croft2018}
{Croft} R.~A.~C., {Miralda-Escud{\'e}} J., {Zheng} Z., {Blomqvist} M., {Pieri}
  M., 2018, \mnras

\bibitem[{{Croft} {et~al}\mbox{.}(2016){Croft}, {Miralda-Escud{\'e}}, {Zheng},
  {Bolton}, {Dawson}, {Peterson}, {York}, \& et~al}]{croft2016}
{Croft} R.~A.~C., {Miralda-Escud{\'e}} J., {Zheng} Z., {Bolton} A., {Dawson}
  K.~S., {Peterson} J.~B., {York} D.~G., et~al, 2016, \mnras, 457, 3541

\bibitem[{{Ellison} {et~al}\mbox{.}(2013){Ellison}, {Mendel}, {Patton}, \&
  {Scudder}}]{Ellison2013}
{Ellison} S.~L., {Mendel} J.~T., {Patton} D.~R., {Scudder} J.~M., 2013, \mnras,
  435, 3627

\bibitem[{{Farage} {et~al}\mbox{.}(2010){Farage}, {McGregor}, {Dopita}, \&
  {Bicknell}}]{Farage2010}
{Farage} C.~L., {McGregor} P.~J., {Dopita} M.~A., {Bicknell} G.~V., 2010, \apj,
  724, 267

\bibitem[{{Gallazzi} {et~al}\mbox{.}(2005){Gallazzi}, {Charlot}, {Brinchmann},
  {White}, \& {Tremonti}}]{Gallazzi}
{Gallazzi} A., {Charlot} S., {Brinchmann} J., {White} S.~D.~M., {Tremonti}
  C.~A., 2005, MNRAS, 362, 41

\bibitem[{{Hani} {et~al}\mbox{.}(2018){Hani}, {Sparre}, {Ellison}, {Torrey}, \&
  {Vogelsberger}}]{Hani2018}
{Hani} M.~H., {Sparre} M., {Ellison} S.~L., {Torrey} P., {Vogelsberger} M.,
  2018, \mnras, 475, 1160

\bibitem[{{Hopkins} \& {Hernquist}(2006)}]{Hopkins2006}
{Hopkins} P.~F., {Hernquist} L., 2006, \apjs, 166, 1

\bibitem[{{Johnson} {et~al}\mbox{.}(2018){Johnson}, {Chen}, {Straka}, {Schaye},
  {Cantalupo}, {Wendt}, {Muzahid}, {Bouch{\'e}}, {Herenz}, {Kollatschny},
  {Mulchaey}, {Marino}, {Maseda}, \& {Wisotzki}}]{Johnson2018}
{Johnson} S.~D. {et~al.}, 2018, \apjl, 869, L1

\bibitem[{{Joshi} {et~al}\mbox{.}(2018){Joshi}, {Srianand}, {Petitjean}, \&
  {Noterdaeme}}]{joshi2018}
{Joshi} R., {Srianand} R., {Petitjean} P., {Noterdaeme} P., 2018, \mnras, 476,
  210

\bibitem[{{Kauffmann} {et~al}\mbox{.}(2003{\natexlab{a}}){Kauffmann},
  {Heckman}, {White}, {Charlot}, {Tremonti}, {Brinchmann}, {Bruzual}, {Peng},
  {Seibert}, {Bernardi}, {Blanton}, {Brinkmann}, {Castander}, {Cs{\'a}bai},
  {Fukugita}, {Ivezic}, {Munn}, {Nichol}, {Padmanabhan}, {Thakar}, {Weinberg},
  \& {York}}]{Kauffmann2003a}
{Kauffmann} G. {et~al.}, 2003{\natexlab{a}}, MNRAS, 341, 33

\bibitem[{{Kauffmann} {et~al}\mbox{.}(2003{\natexlab{b}}){Kauffmann},
  {Heckman}, {White}, {Charlot}, {Tremonti}, {Peng}, {Seibert}, \&
  et~al}]{Kauffmann2003b}
{Kauffmann} G., {Heckman} T.~M., {White} S.~D.~M., {Charlot} S., {Tremonti} C.,
  {Peng} E.~W., {Seibert} M., et~al, 2003{\natexlab{b}}, \mnras, 341, 54

\bibitem[{{Kere{\v s}} {et~al}\mbox{.}(2005){Kere{\v s}}, {Katz}, {Weinberg},
  \& {Dav{\'e}}}]{keres}
{Kere{\v s}} D., {Katz} N., {Weinberg} D.~H., {Dav{\'e}} R., 2005, \mnras, 363,
  2

\bibitem[{{Kewley}, {Geller} \& {Barton}(2006){Kewley}, {Geller}, \&
  {Barton}}]{Kewley2006}
{Kewley} L.~J., {Geller} M.~J., {Barton} E.~J., 2006, \aj, 131, 2004

\bibitem[{{Kewley} {et~al}\mbox{.}(2010){Kewley}, {Rupke}, {Zahid}, {Geller},
  \& {Barton}}]{Kewley2010}
{Kewley} L.~J., {Rupke} D., {Zahid} H.~J., {Geller} M.~J., {Barton} E.~J.,
  2010, \apjl, 721, L48

\bibitem[{{Klypin} {et~al}\mbox{.}(2016){Klypin}, {Yepes}, {Gottl{\"o}ber},
  {Prada}, \& {He{\ss}}}]{Klypin16}
{Klypin} A., {Yepes} G., {Gottl{\"o}ber} S., {Prada} F., {He{\ss}} S., 2016,
  \mnras, 457, 4340

\bibitem[{{Lan} \& {Mo}(2018)}]{lan2018}
{Lan} T.-W., {Mo} H., 2018, ArXiv 1806.05786

\bibitem[{{Malin} \& {Hadley}(1997)}]{malin}
{Malin} D., {Hadley} B., 1997, \pasa, 14, 52

\bibitem[{{M{\'e}nard} {et~al}\mbox{.}(2011){M{\'e}nard}, {Wild}, {Nestor},
  {Quider}, {Zibetti}, {Rao}, \& {Turnshek}}]{menard2011}
{M{\'e}nard} B., {Wild} V., {Nestor} D., {Quider} A., {Zibetti} S., {Rao} S.,
  {Turnshek} D., 2011, \mnras, 417, 801

\bibitem[{{Nardini} {et~al}\mbox{.}(2013){Nardini}, {Wang}, {Fabbiano},
  {Elvis}, {Pellegrini}, {Risaliti}, {Karovska}, \& {Zezas}}]{Nardini2013}
{Nardini} E., {Wang} J., {Fabbiano} G., {Elvis} M., {Pellegrini} S., {Risaliti}
  G., {Karovska} M., {Zezas} A., 2013, \apj, 765, 141

\bibitem[{{Pettini} \& {Pagel}(2004)}]{Pettini2004}
{Pettini} M., {Pagel} B. E.~J., 2004, \mnras, 348, L59

\bibitem[{{Planck Collaboration} {et~al}\mbox{.}(2018){Planck Collaboration},
  {Akrami}, {Arroja}, {Ashdown}, {Aumont}, {Baccigalupi}, {Ballardini},
  {Banday}, {Barreiro}, \& {Bartolo}}]{Planck2018}
{Planck Collaboration} {et~al.}, 2018, arXiv e-prints, arXiv:1807.06205

\bibitem[{{Prochaska} {et~al}\mbox{.}(2017){Prochaska}, {Werk}, {Worseck},
  {Tripp}, {Tumlinson}, \& et~al}]{prochaska2017}
{Prochaska} J.~X., {Werk} J.~K., {Worseck} G., {Tripp} T.~M., {Tumlinson} J.,
  et~al, 2017, \apj, 837, 169

\bibitem[{{Rampazzo} {et~al}\mbox{.}(2005){Rampazzo}, {Plana}, {Amram},
  {Bagarotto}, {Boulesteix}, \& {Rosado}}]{Rampazzo2005}
{Rampazzo} R., {Plana} H., {Amram} P., {Bagarotto} S., {Boulesteix} J.,
  {Rosado} M., 2005, \mnras, 356, 1177

\bibitem[{{Rich} {et~al}\mbox{.}(2010){Rich}, {Dopita}, {Kewley}, \&
  {Rupke}}]{Rich2010}
{Rich} J.~A., {Dopita} M.~A., {Kewley} L.~J., {Rupke} D.~S.~N., 2010, \apj,
  721, 505

\bibitem[{{Riess} {et~al}\mbox{.}(2018){Riess}, {Casertano}, {Yuan}, {Macri},
  {Bucciarelli}, {Lattanzi}, {MacKenty}, {Bowers}, {Zheng}, {Filippenko},
  {Huang}, \& {Anderson}}]{riess}
{Riess} A.~G. {et~al.}, 2018, \apj, 861, 126

\bibitem[{{Rodr{\'{\i}}guez-Puebla}
  {et~al}\mbox{.}(2016){Rodr{\'{\i}}guez-Puebla}, {Behroozi}, {Primack},
  {Klypin}, {Lee}, \& {Hellinger}}]{RP16}
{Rodr{\'{\i}}guez-Puebla} A., {Behroozi} P., {Primack} J., {Klypin} A., {Lee}
  C., {Hellinger} D., 2016, \mnras, 462, 893

\bibitem[{{Simard} {et~al}\mbox{.}(2011){Simard}, {Mendel}, {Patton},
  {Ellison}, \& {McConnachie}}]{simard}
{Simard} L., {Mendel} J.~T., {Patton} D.~R., {Ellison} S.~L., {McConnachie}
  A.~W., 2011, ApJs, 196, 11

\bibitem[{{Spitzer}(1956)}]{spitzer}
{Spitzer}, Jr. L., 1956, \apj, 124, 20

\bibitem[{{Springel}, {Di Matteo} \& {Hernquist}(2005){Springel}, {Di Matteo},
  \& {Hernquist}}]{Springel2005}
{Springel} V., {Di Matteo} T., {Hernquist} L., 2005, \apjl, 620, L79

\bibitem[{{Steidel} {et~al}\mbox{.}(2010){Steidel}, {Erb}, {Shapley},
  {Pettini}, {Reddy}, {Bogosavljevi{\'c}}, {Rudie}, \& {Rakic}}]{steidel2010}
{Steidel} C.~C., {Erb} D.~K., {Shapley} A.~E., {Pettini} M., {Reddy} N.,
  {Bogosavljevi{\'c}} M., {Rudie} G.~C., {Rakic} O., 2010, \apj, 717, 289

\bibitem[{{Thilker} {et~al}\mbox{.}(2005){Thilker}, {Bianchi}, {Boissier}, {Gil
  de Paz}, {Madore}, {Martin}, {Meurer}, {Neff}, {Rich}, {Schiminovich},
  {Seibert}, {Wyder}, {Barlow}, {Byun}, {Donas}, {Forster}, {Friedman},
  {Heckman}, {Jelinsky}, {Lee}, {Malina}, {Milliard}, {Morrissey}, {Siegmund},
  {Small}, {Szalay}, \& {Welsh}}]{thilker}
{Thilker} D.~A. {et~al.}, 2005, \apjl, 619, L79

\bibitem[{{Toomre} \& {Toomre}(1972)}]{toomre}
{Toomre} A., {Toomre} J., 1972, \apj, 178, 623

\bibitem[{{Tumlinson}, {Peeples} \& {Werk}(2017){Tumlinson}, {Peeples}, \&
  {Werk}}]{CGM2017}
{Tumlinson} J., {Peeples} M.~S., {Werk} J.~K., 2017, \araa, 55, 389

\bibitem[{{Tytler}(1982)}]{tytler}
{Tytler} D., 1982, \nat, 298, 427

\bibitem[{{Werk} {et~al}\mbox{.}(2016){Werk}, {Prochaska}, {Cantalupo}, {Fox},
  {Oppenheimer}, {Tumlinson}, {Tripp}, \& et~al}]{werk16}
{Werk} J.~K., {Prochaska} J.~X., {Cantalupo} S., {Fox} A.~J., {Oppenheimer} B.,
  {Tumlinson} J., {Tripp} T.~M., et~al, 2016, \apj, 833, 54

\bibitem[{{Werk} {et~al}\mbox{.}(2014){Werk}, {Prochaska}, {Tumlinson},
  {Peeples}, {Tripp}, {Fox}, \& et~al}]{Werk2014}
{Werk} J.~K., {Prochaska} J.~X., {Tumlinson} J., {Peeples} M.~S., {Tripp}
  T.~M., {Fox} A.~J., et~al, 2014, \apj, 792, 8

\bibitem[{{Werk} {et~al}\mbox{.}(2010){Werk}, {Putman}, {Meurer}, {Ryan-Weber},
  {Kehrig}, {Thilker}, {Bland -Hawthorn}, {Drinkwater}, {Kennicutt}, {Wong},
  {Freeman}, {Oey}, {Dopita}, {Doyle}, {Ferguson}, {Hanish}, {Heckman},
  {Kilborn}, {Kim}, {Knezek}, {Koribalski}, {Meyer}, {Smith}, \&
  {Zwaan}}]{werk10}
{Werk} J.~K. {et~al.}, 2010, \aj, 139, 279

\bibitem[{{Yoshida} {et~al}\mbox{.}(2016){Yoshida}, {Yagi}, {Ohyama},
  {Komiyama}, {Kashikawa}, {Tanaka}, \& {Okamura}}]{Yoshida2016}
{Yoshida} M., {Yagi} M., {Ohyama} Y., {Komiyama} Y., {Kashikawa} N., {Tanaka}
  H., {Okamura} S., 2016, \apj, 820, 48

\bibitem[{{Zabludoff} {et~al}\mbox{.}(1996){Zabludoff}, {Zaritsky}, {Lin},
  {Tucker}, {Hashimoto}, {Shectman}, {Oemler}, \& {Kirshner}}]{zabludoff}
{Zabludoff} A.~I., {Zaritsky} D., {Lin} H., {Tucker} D., {Hashimoto} Y.,
  {Shectman} S.~A., {Oemler} A., {Kirshner} R.~P., 1996, \apj, 466, 104

\bibitem[{{Zahedy} {et~al}\mbox{.}(2019){Zahedy}, {Chen}, {Johnson}, {Pierce},
  {Rauch}, {Huang}, {Weiner}, \& {Gauthier}}]{Zahedy2019}
{Zahedy} F.~S., {Chen} H.-W., {Johnson} S.~D., {Pierce} R.~M., {Rauch} M.,
  {Huang} Y.-H., {Weiner} B.~J., {Gauthier} J.-R., 2019, \mnras, 484, 2257

\bibitem[{{Zahedy} {et~al}\mbox{.}(2017){Zahedy}, {Chen}, {Rauch}, \&
  {Zabludoff}}]{Zahedy2017}
{Zahedy} F.~S., {Chen} H.-W., {Rauch} M., {Zabludoff} A., 2017, \apjl, 846, L29

\bibitem[{{Zhang} {et~al}\mbox{.}(2020){Zhang}, {Yang}, {Zaritsky}, {Behroozi},
  \& {Werk}}]{Zhang2020}
{Zhang} H., {Yang} X., {Zaritsky} D., {Behroozi} P., {Werk} J., 2020, arXiv
  e-prints, arXiv:1911.02032

\bibitem[{{Zhang}, {Zaritsky} \& {Behroozi}(2018){Zhang}, {Zaritsky}, \&
  {Behroozi}}]{zhang2018a}
{Zhang} H., {Zaritsky} D., {Behroozi} P., 2018, \apj, 861, 34

\bibitem[{{Zhang} {et~al}\mbox{.}(2019){Zhang}, {Zaritsky}, {Behroozi}, \&
  {Werk}}]{Zhang2019a}
{Zhang} H., {Zaritsky} D., {Behroozi} P., {Werk} J., 2019, \apj, 880, 28

\bibitem[{{Zhang} {et~al}\mbox{.}(2018){Zhang}, {Zaritsky}, {Werk}, \&
  {Behroozi}}]{Zhang2018b}
{Zhang} H., {Zaritsky} D., {Werk} J., {Behroozi} P., 2018, \apj, 866, L4

\bibitem[{{Zhang} {et~al}\mbox{.}(2016){Zhang}, {Zaritsky}, {Zhu},
  {M{\'e}nard}, \& {Hogg}}]{zhang2016}
{Zhang} H., {Zaritsky} D., {Zhu} G., {M{\'e}nard} B., {Hogg} D.~W., 2016, \apj,
  833, 276

\bibitem[{{Zhu} \& {M{\'e}nard}(2013{\natexlab{a}})}]{zhu2013a}
{Zhu} G., {M{\'e}nard} B., 2013{\natexlab{a}}, \apj, 773, 16

\bibitem[{{Zhu} \& {M{\'e}nard}(2013{\natexlab{b}})}]{zhu2013b}
{Zhu} G., {M{\'e}nard} B., 2013{\natexlab{b}}, \apj, 770, 130

\end{thebibliography}

\end{document}